\documentclass[preprint,12pt]{elsarticle}

\usepackage{amssymb}

\usepackage{xcolor}
\usepackage{subfigure}
\usepackage{float}
\usepackage{adjustbox}
\usepackage{placeins}
\usepackage[T1]{fontenc}
\usepackage{csquotes}

\usepackage[nointegrals]{wasysym}
\usepackage{amsmath}
\DeclareMathOperator{\sech}{sech}
\usepackage{amssymb}
\usepackage{amsthm}
\usepackage{braket}
\usepackage{bm}
\usepackage{mathtools}
\usepackage{tensor}
\usepackage{mathrsfs}

\def\openone{\leavevmode\hbox{\small$1$\normalsize\kern-.33em$1$}}

\newcommand{\abs}[1]{\left| #1 \right|} 
\newcommand{\avg}[1]{\left< #1 \right>}

\let\baraccent=\= 
\renewcommand{\=}[1]{\stackrel{#1}{=}}

\newcommand{\be}{\begin{equation}}
\newcommand{\bel}[1]{\begin{equation}\label{#1}}
\newcommand{\ee}{\end{equation}}

\newcommand{\ii}{\mathrm{i}}

\newcommand{\dd}{\mathrm{d}}

\journal{Journal of Magnetic Resonance}

\begin{document}

\begin{frontmatter}



\title{Frame Change Technique for Phase Transient Cancellation}


\author[MITNSE,MITRLE]{Andrew Stasiuk}
\author[Princeton]{Pai Peng}
\author[MITPhys]{Garrett Heller}
\author[MITNSE,MITRLE,MITPhys]{Paola Cappellaro}

\affiliation[MITNSE]{organization={Department of Nuclear Science and Engineering},
            addressline={Massachusetts Institute of Technology}, 
            city={Cambridge},
            postcode={02139}, 
            state={MA},
            country={USA}}

\affiliation[MITRLE]{organization={Research Laboratory of Electronics},
            addressline={Massachusetts Institute of Technology}, 
            city={Cambridge},
            postcode={02139}, 
            state={MA},
            country={USA}}

\affiliation[Princeton]{organization={Department of Electrical and Computer Engineering},
            addressline={Princeton University}, 
            city={Princeton},
            postcode={08544}, 
            state={NJ},
            country={USA}}


\affiliation[MITPhys]{organization={Department of Physics},
            addressline={Massachusetts Institute of Technology}, 
            city={Cambridge},
            postcode={02139}, 
            state={MA},
            country={USA}}

\begin{abstract}
The precise control of complex quantum mechanical systems can unlock applications ranging from quantum simulation to quantum computation. Controlling strongly interacting many-body systems often relies on Floquet Hamiltonian engineering that is achieved by fast switching between Hamiltonian primitives via external control. For example, in our solid-state NMR system, we perform quantum simulation by modulating the natural Hamiltonian with control pulses. As the Floquet heating errors scale with the interpulse delay, $\delta t$, it is favorable to keep $\delta t$ as short as possible, forcing our control pulses to be short duration and high power. Additionally, high-power pulses help to minimize undesirable evolution from occurring during the duration of the pulse.  However, such pulses introduce an appreciable phase-transient control error, a form of unitary error. In this work, we detail our ability to diagnose the error, calibrate its magnitude, and correct it for $\pi/2$-pulses of arbitrary phase. We demonstrate the improvements gained by correcting for the phase transient error, using a method which we call the ``frame-change technique'', in a variety of experimental settings of interest. Given that the correction mechanism adds no real control overhead, we recommend that any resonance probe be checked for these phase transient control errors, and correct them using the frame-change technique.
\end{abstract}

\begin{keyword}
nuclear magnetic resonance \sep solid-state NMR \sep fluorapatite \sep high-fidelity control \sep pulse error cancellation
\end{keyword}

\end{frontmatter}


\section{Introduction}\label{sec:intro}
As a fairly mature technology, nuclear magnetic resonance (NMR) provides easy access to systems which evolve under quantum mechanical equations of motion. In particular for solid-state systems, multiple-pulse experiments were introduced to increase the spectral resolution, thereby extracting more information about a sample \cite{waugh1968approach,haeberlen1976high}. Additionally, solid-state NMR provides access to strongly interacting quantum systems with $T_1$ times on the order of seconds, which is of interest for condensed matter theory. The accessible dynamics of these systems are greatly expanded via the Hamiltonian engineering toolbox, wherein periodic application of pulse sequence gives rise to average evolution under a desirable target Hamiltonian \cite{mananga2016floquet,mansfield1971symmetrized,peng2022deep,maricq1982application}. Proper utilization of these techniques situates solid-state NMR as a leading method for analog quantum simulation. Of vital importance is ensuring that the simulated dynamics does not differ too much from the target dynamics. Achieving high-fidelity simulations and measuring high-resolution spectra requires good hardware, a well calibrated device, and well designed pulse sequences \cite{heyl2019trotter,sieberer2019digital}. 

As early as the 1970s, it was realized that attempts to produce square pulses for the purpose of radio-frequency quantum control led to electronic ``phase glitches'', more commonly known as phase transients, as a result of non-linear circuit elements \cite{mehring1972phase}. Namely during the rise and fall time of an RLC circuit under a square pulse voltage driving, there is a naturally occurring out-of-phase contribution to the magnetic field experienced by the spins in the rotating frame \cite{ellett1971spectrometers,haeberlen1976high}. Investigations of this effect were later extended from simple RLC circuits to parallel-tuned, series-matched probe circuits to better model NMR systems \cite{barbara1991phase}. It has been repeatedly found, and indeed it is expected, that the actual phase transient is larger in magnitude than the electronic analysis would otherwise predict \cite{ellett1971spectrometers,barbara1991phase}.

Early solutions to correcting for the effects of phase transients were predicated on the symmetry of the magnitude of the leading and trailing edge errors. In these older systems, the relative magnitude of these leading and trailing terms could be controlled via modification of the quiescent base currents of transistors within the probe electronics \cite{haeberlen1976high}. Even so, special multi-pulse sequences were designed to be robust to phase-transients, often extending the Floquet period of a given cycle, such as XY-4,8,16, which are self-correcting versions of CPMG \cite{carr1954effects,maudsley1986modified,souza2012robust}. Indeed, it would seem that modern spectrometers have less control over the symmetry of the leading and trailing phase transients \cite{wittmann2015compensating}. Via pulse shaping, the phase transient can be nearly entirely limited, especially with the help of optimal control theory techniques, at the cost of longer pulse lengths \cite{wittmann2015compensating,hincks2015controlling}.

When the experimental goal is to shape the system dynamics via Hamiltonian engineering, such longer pulse lengths are usually detrimental and increase the error in generating a desired time-average target Hamiltonian. To minimize the Trotter error during an evolution of length $T = n \delta t$, one takes each Trotter step $\delta t$ to be as short as possible. This in turn implies that the delays within a Floquet cycle designed to produce a target Hamiltonian should also be as short as possible. The physical limitations that prevent $\delta t \longrightarrow 0$ are finite pulse lengths and electronic switching times. By taking $\delta t$ to be as short as possible, we introduce significant phase transient errors into the control, but we still find that longer, albeit higher fidelity, pulses perform worse than short high-power pulses 

Here we show how to overcome this challenge by canceling the phase transient error for $\pi/2$-pulses entirely  with the inclusion of virtual-z gates, which allow for a high fidelity rotations with 0 pulse length \cite{mckay2017efficient,knill2000algorithmic}. These virtual-z gates have been utilized in NMR to introduce static fields \cite{wei2018localization} and in superconducting quantum platforms to cancel off-resonant control errors \cite{mckay2017efficient}. Inspired by their usage in superconducting devices, we calibrate our unitary pulse error stemming from the phase transient, and are able to fully correct it using only virtual-z rotations. In this work we detail how to calibrate and correct the phase transient error, and demonstrate significant improvements in Hamiltonian engineering and dynamical decoupling fidelity.

\section{Methods}

\begin{figure}[t!]
    \centering
    \includegraphics[width=\textwidth]{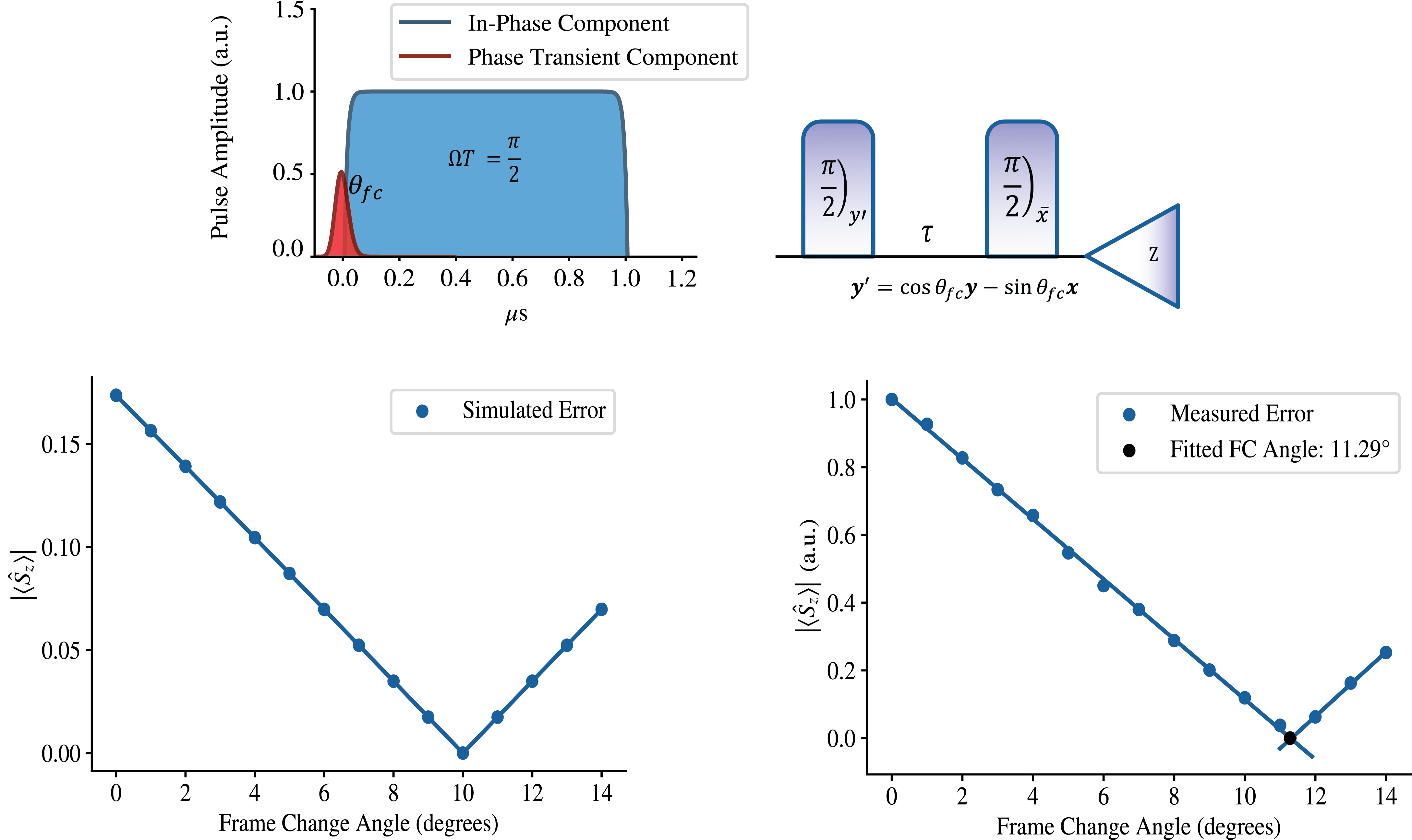}
    \caption{Schematic of the phase transient introduced by short square pulses (top-left), and the experimental sequence to determine $\theta_{fc}$ (top-right). 
    If there were no phase transient error, then we would expect to measure $\avg{\hat{S}_z}=0$. By varying the deflection angle from the $y$-axis of the first pulse, we can experimentally determine $\theta_{fc}$ by finding the angle which minimizes $\avg{\hat{S}_z}$. In the bottom-left, we show the theoretical prediction for this error model when $\theta_{fc} = 10^{\circ}$, along with experimental data (bottom-right) which demonstrates a measured $\theta_{fc}\approx11^{\circ}$.} 
    \label{fig1}
\end{figure}

We focus on controlling an ensemble of solid-state nuclear spins via on-resonance radio-frequency (rf) pulses resulting in $\pi/2$ pulses. We perform experiments on a single crystal of fluorapatite, Ca$_5$(PO$_4$)$_3$F, in a 300 MHz ($\sim 7$T) Bruker spectrometer. The crystal is placed with the magnetic field aligned along its $c$ axis to maximize nearest-neighbor fluorine-fluorine coupling strength. Ideally, our control induces the following Hamiltonian in the rotating frame,
\begin{equation}
    \mathcal{H}_c(t) = f(t) \big(\cos(\phi) \hat{S}_x + \sin(\phi)\hat{S}_y\big),
\end{equation}
for collective spin operators $\hat{S}_\mu$, $\mu\in\lbrace x, y, z\rbrace$. The function $f(t)$ is ideally a train of square pulses of fixed length and power, each resulting in a net $\pi/2$ rotation about the transverse axis determined by $\phi$, which has a resolution of one degree. 

In reality, true square pulses are impossible to generate, as there is a finite rise and fall time when turning on and off a voltage source. It was found that during the leading and trailing tails, an out-of-phase current component is induced in the resonance probe circuit \cite{mehring1972phase, barbara1991phase}. The magnitude and relative ratio of these leading and trailing phase transients is affected by the pulse power and the quiescent base current of the control circuit \cite{mehring1972phase}. The latter indicator, the quiescent base current, is not readily modifiable on our spectrometer, or most modern spectrometers for that matter \cite{wittmann2015compensating}.

Indeed, we find that for short $\pi/2$ pulses, $t_p < 2.5\mu$s, there is a noticeable phase transient contribution, dominantly seen in the leading edge (see Fig.~\ref{fig1} and \ref{sec.app_pulse}). Namely, for each applied $\pi/2$ pulse, there is an additional small angle rotation with consistent amplitude, $\theta_{fc}$, 90 degrees out of phase. That is,
\begin{equation}\label{eqn.error_model}
    U_\phi = R_\phi(\pi/2) \longrightarrow U_{exp} = R_\phi(\pi/2)R_{\phi+\pi/2}(\theta_{fc}),
\end{equation}
where
\begin{equation}
    R_\phi(\theta) = \exp\bigg(-\ii \theta \big(\cos\phi \hat{S}_x + \sin\phi \hat{S}_y\big)\bigg).
\end{equation}

In accordance with predictions from the electronics, we have found that $\theta_{fc}$ is fixed for a given tune-up setting. Namely, for a given tuning and matching of the resonance probe, we calibrate the $\pi/2$ power which then fixes the frame-change angle. This angle is stable until the probe needs to be re-tuned, usually in accordance with large external temperature and humidity fluctuations. We now demonstrate that the value of $\theta_{fc}$ can be efficiently experimentally determined. This procedure is detailed in Figure \ref{fig1}. Once $\theta_{fc}$ is calibrated, the error can be corrected by noting the following identity:
\begin{equation}
   R_z(\mp\theta_{fc}) R_\phi(\pi/2)  R_{\phi\pm\pi/2}(\theta_{fc})= R_\phi(\pi/2).
\end{equation}
Thus, the out-of-phase over-rotation error can be completely canceled via a repeated $z$ rotation after each electronically generated pulse.

\begin{figure}
    \centering
    \includegraphics[scale=.85]{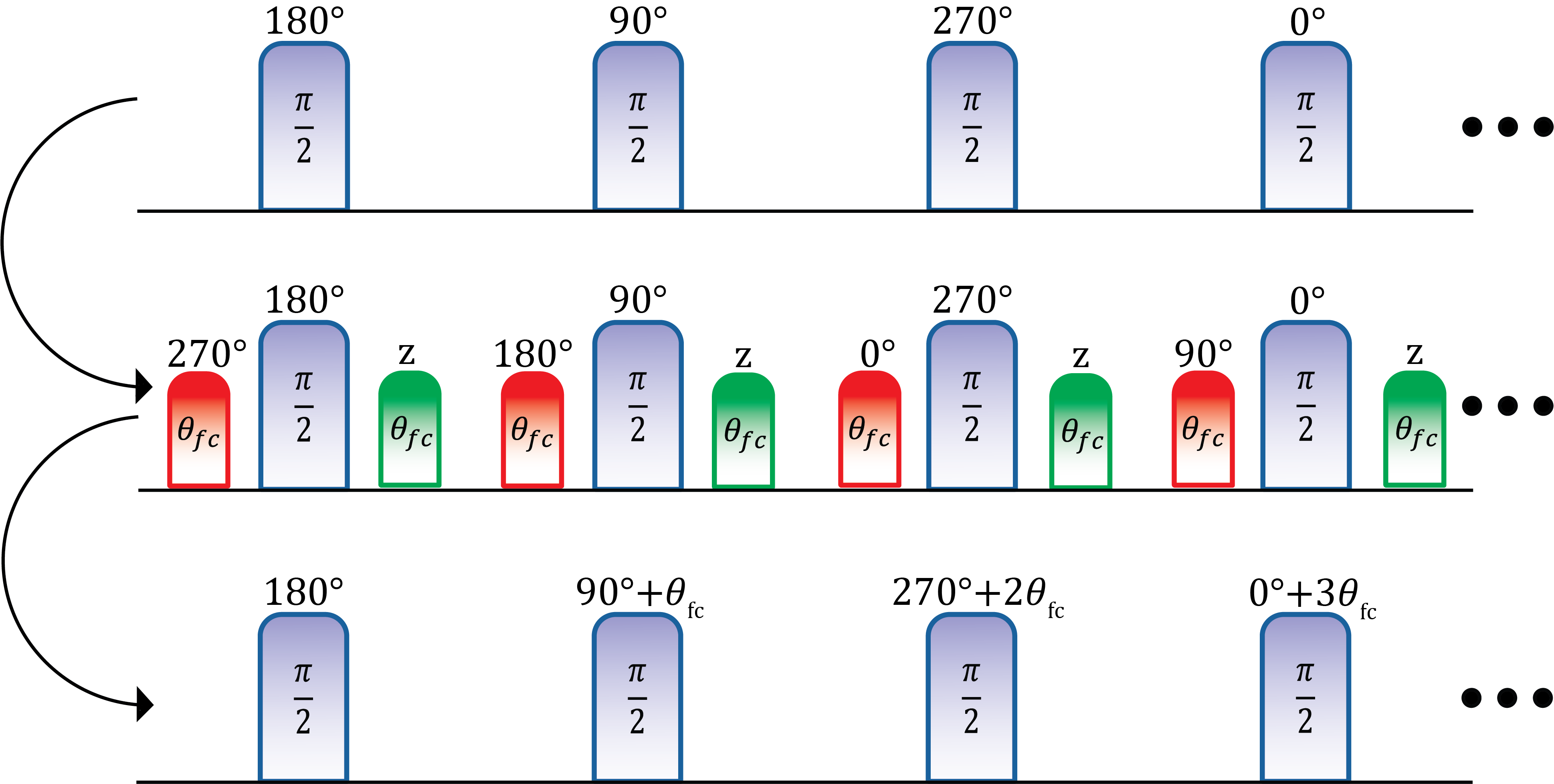}
    \caption{Schematic representation of the frame-change technique, applied to a WAHUHA-like pulse sequence. On the top, we show the \textit{logical pulse sequence}, a series of $\pi/2$ pulses of arbitrary phases. We then include the phase transient error (red), and the necessary frame-change correction $z$-rotations (green), to form the \textit{detailed pulse sequence} (middle). Finally, we compile the $z$-rotations, which implicitly cancel the errors, resulting in the \textit{experimental pulse sequence} (bottom), which implements the logical pulse sequence and cancels the phase transient errors.}
    \label{fig2}
\end{figure}

Since a $z$-rotation of angle $\theta$ is performed by phase shifting all subsequent pulses by $\theta$, our method for error correction is achieved by an increasing phase shift, whereby the $n$-th pulse is shifted by an angle $(n-1)\theta_{fc}$ from its original value. This is pictorially shown in Figure \ref{fig2}. Our method can be thought of as a discrete version of a rotating frame transformation, and we call $\theta_{fc}$ the ``frame-change'' angle. In the next section, we will demonstrate the improvements gained from correcting for this error.

Finally, we note that this technique still works, albeit approximately, if the phase transient error is balanced over the leading and trailing edge. It has previously been shown that arbitrary phase transient error results in a slight deviation of the rotation axis into the $z$-direction, resulting in the impossibility of performing a $\pi$-rotation in a single pulse \cite{ellett1971spectrometers}. Concretely, for a desired $\pi/2$ $x$-rotation, the experimental unitary is modified as
\begin{equation}
   R_x(\pi/2) \longrightarrow U_{alt} = \exp\bigg(-\ii\big(\frac{\pi}{2}\hat{S}_x + \epsilon_{fc} \hat{S}_z\big)\bigg).
\end{equation}
We see that the phase transient error here manifests as a deflection of the rotation axis (by some angle $\epsilon_{fc}$) which can also be canceled to leading order with the frame-change method. Given that the electronics of our system are leading-edge dominated, we focus on this case, where the error can be cancelled exactly to within one degree.

\section{Results and Discussion}

We evaluate the performance of the frame-change technique by measuring  improved fidelity in Hamiltonian engineering and longer-lived signals in dynamical decoupling experiments. Indeed, across a wide range of experimental settings, by correcting the out-of-phase over-rotation error the frame-change technique leads to improved signal lifetimes. We demonstrate this effect in time-suspension experiments, Loschmidt echo (time reversal) experiments, and spectral sequence experiments.

In addition, we perform these experiments for multiple pulse lengths. Generally, the shorter the pulse length, the higher the required power, which in turn results in a larger $\theta_{fc}$. This trend is consistent with the existing literature's conclusions on the electronic origins of the phase transient \cite{haeberlen1976high}. However, shorter pulses also allow for shorter delays during Floquet engineering, leading to smaller errors in simulating a target Hamiltonian \cite{sieberer2019digital,heyl2019trotter}. We will clearly demonstrate that cancellation of the phase transient error via the frame-change technique allows for higher resolution experiments at equivalent pulse lengths. However, it will also be clear that minimizing the Trotter error has a greater impact on achieving high-fidelity evolution under a target time-averaged Hamiltonian. Hence, short, high-power pulses using the frame-change technique offer the highest performance possible on our hardware.

\begin{figure}[t!]
    \centering
    \includegraphics[scale=0.8]{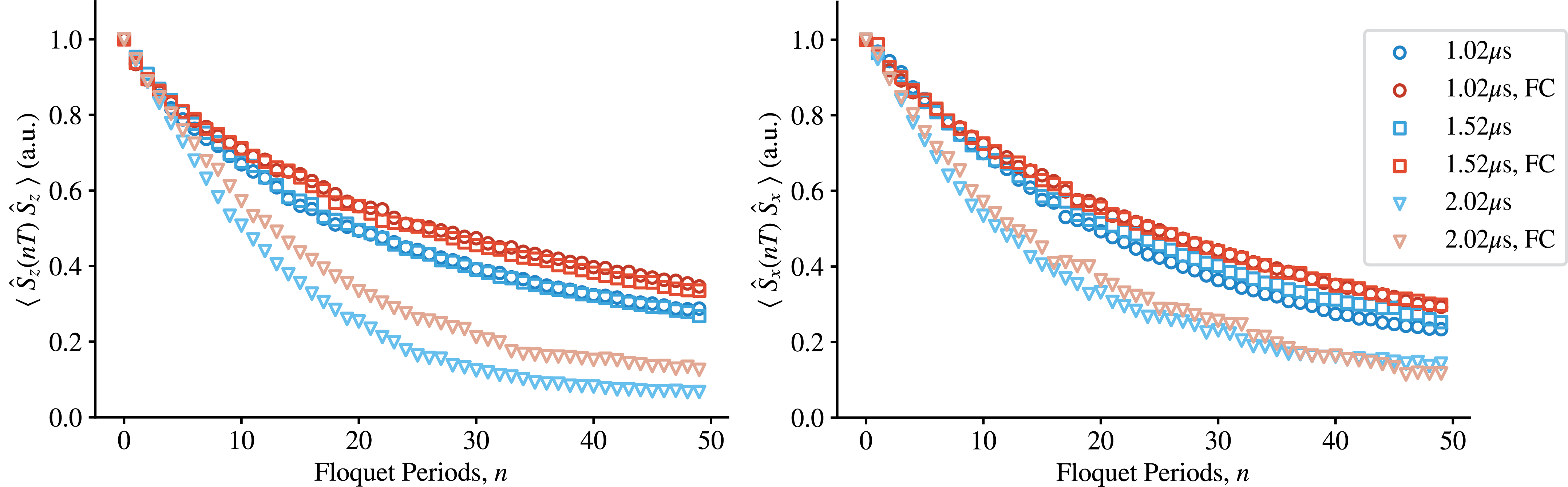}
    \caption{Comparison of the fidelity of Peng24 time suspension sequences with (red) and without (blue) frame-change correction in both the longitudinal (left) and transverse (right) directions. For pulse lengths of $1.02 \mu$s and $1.52 \mu$s, we use the delay primitive $\tau = 5 \mu$s, leading to a Floquet period of $120 \mu$s. For $t_p=2.02 \mu$s, the delay primitive must be increased to $\tau = 6\mu$s, leading to a larger error term manifesting in a signal of significantly reduced magnitude.}
    \label{fig3}
\end{figure}

In Figure \ref{fig3}, we demonstrate that under the Peng24 time suspension sequence\cite{peng2022deep}, the frame-change correction leads to a slower decay, for various pulse lengths $t_p = 1.02, 1.52, 2.02 \mu$s. Peng24 was specifically designed to be robust to unitary errors, so one might think that the frame-change technique would not show a significant improvement \cite{peng2022deep}. However, the phase transient-induced error is markedly different than a simple over-rotation error, and we find that correcting for this error can still provide meaningful improvements in the signal lifetime. Even if a pulse sequence was designed to cancel the phase-transient error over its Floquet period, such cancellation would likely still be imperfect, and thus benefit from the addition of the frame-change technique. 

\begin{figure}[t]
    \centering
    \includegraphics[scale=0.85]{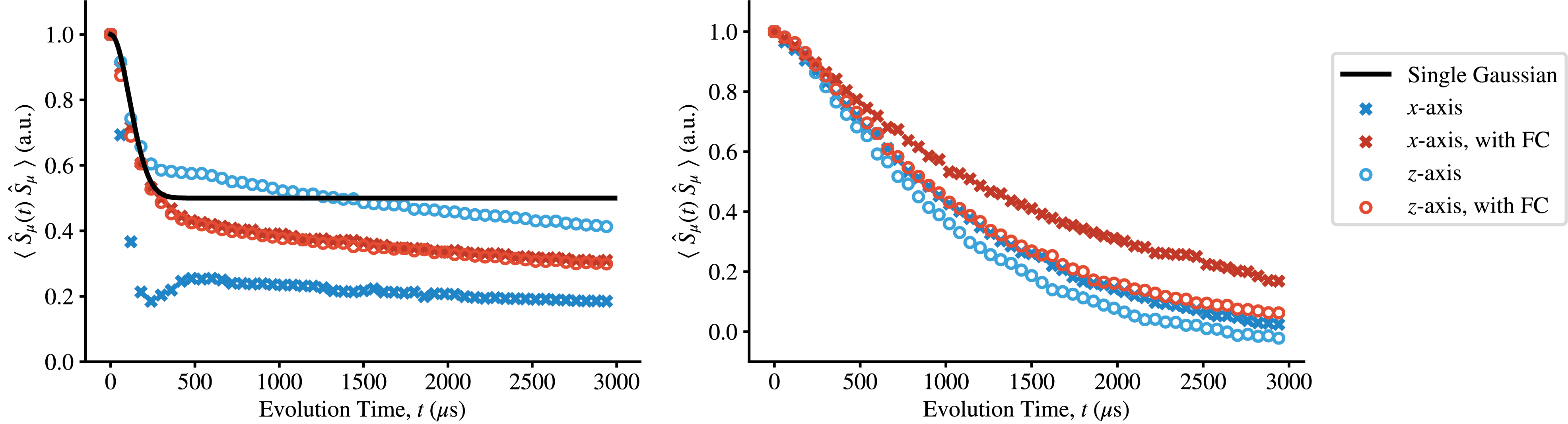}
    \caption{ On the left, we show a comparison of analog quantum simulation results with (red) and without (blue) the frame-change correction for pulse length $t_p = 1.02\mu$s under MREV-8. Additionally, we include the analytic prediction (black), showing good coincidence with the frame-changed corrected experimental data. On the right, we show the evolution under the Angle12 time suspension sequence, which has poorer Trotter error scaling than Peng24, leading to a more pronounced difference than shown in Figure \ref{fig3}.}
    \label{fig4}
\end{figure}

Generally, including the frame-change technique should not reduce the signal fidelity. However, the magnitude of improvement can be variable depending on the sequence and the quadrature of measurement. Indeed, this effect is even more pronounced if we consider Angle12, a time suspension sequence consisting of only the first half of Peng24 \cite{peng2022deep}. This sequence has no reflection symmetry, and thus \textit{a priori} one would expect it to have worse error scaling than Peng24. The results of this experiment are shown in Figure \ref{fig4}.

Next, we analyze a case in which including frame-change corrections results in a decrease in signal magnitude, but an increase in the simulation fidelity. For this, we utilize the MREV-8 spectral sequence, which cancels the dipolar interaction, but leaves a residual disordered field term \cite{haeberlen1968coherent,mansfield1971symmetrized}. We choose MREV-8 over other spectral sequences, such as symmetrized WAHUHA, because the resulting disordered field points along the $\frac{1}{\sqrt{2}}(\bm{x} + \bm{z})$ axis (with scaling factor $\sqrt{2}/3$). Thus, we \textit{a priori} expect measurements in the $x$ and $z$ quadrature to be equal. Indeed, under the assumption of quenched disorder, with $h_i$ distributed as a Gaussian with 0 mean and variance $\sigma_h^2$, we compute
\begin{equation}
    \overline{\avg{\hat{S}_z(t)\hat{S}_z}} = \overline{\avg{\hat{S}_x(t)\hat{S}_x}} = \frac{1}{2} + \frac{1}{2}e^{-\frac{1}{9}(\sigma_h t)^2}.
\end{equation}
Details of this calculation are given in \ref{sec.app_mrev}. In figure \ref{fig4}, we show the results of these measurements, with and without frame change corrections, along with the analytic prediction.

To understand the over-prediction of the $z$-measurements, we recall that the net effect of a general phase transient error, integrated over the duration of the pulse, is a deflection of the rotation axis towards the $z$-direction, with the sign of the deflection depending on the details of the phase transient \cite{ellett1971spectrometers}. Hence, to leading order, this error induces a Trotterized field along $z$ over each Floquet period. In total, the effective field felt by each spin is
\begin{equation}
    \bm{B}_\epsilon^{(i)} = \frac{1}{3}h_i \bm{x} + \frac{1}{3}\big(h_i + 3 h_\epsilon\big)\bm{z},
\end{equation}
where $h_\epsilon$ is the magnitude of the field generated by the phase transient error over a single Floquet period. This implies that magnetization along $z$ is ``more conserved'' than it otherwise should be, thanks to an emerging symmetry. Conversely, the $x$ magnetization is not protected (its fast decay prevents us from observing oscillations due to $h_\epsilon$). These measurements conclusively show that the frame change technique is capable of significantly improving the simulation fidelity, as evidenced by the coincidence of $\overline{\avg{\hat{S}_z(t)\hat{S}_z}}$ and $\overline{\avg{\hat{S}_x(t)\hat{S}_x}}$ with frame change corrections.

\begin{figure}
    \centering
    \includegraphics[scale=0.85]{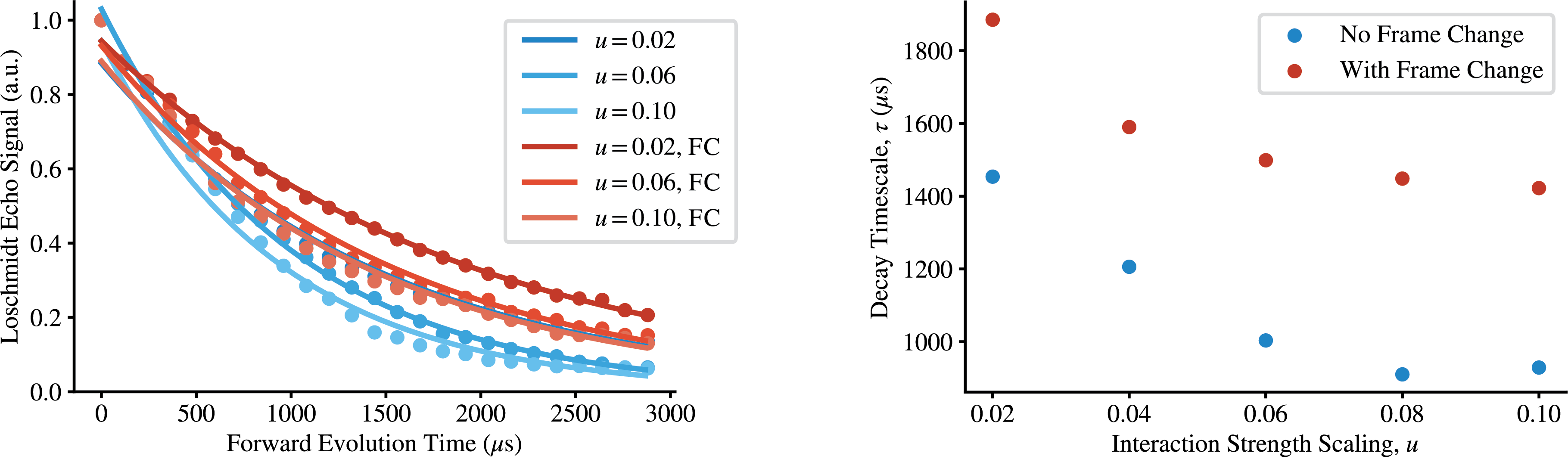}
    \caption{Effect of the frame change technique on a Loschmidt echo experiment. On the left, we show the signal as a function of the forward evolution time for select interaction scalings, $u$. Experiments (not) using the frame change technique are plotted in red (blue), along with their fits to the model $S(t) = A\exp(-(t/\tau))$. On the right, we plot the decay timescale $\tau$ as a function of the interaction scaling parameter $u$. Notice that the inclusion of the frame change technique leads to the Loschmidt echo to decaying more slowly than the results from evolving under the uncorrected sequence.}
    \label{fig5}
\end{figure}

In addition to time suspension and dipolar decoupling sequences, we analyze Hamiltonian engineering pulse sequences which result in a non-zero target Hamiltonian. To evaluate the control performance in this case, we measure the Loschmidt echo. The Loschmidt echo is a fidelity-like measurement linked to irreversibility and quantum chaos, which encodes our ability to reverse Hamiltonian evolution \cite{gorin2006dynamics}. For this investigation, we use Wei16 to engineer the double quantum Hamiltonian for a variety of interaction strengths, and measure the Loschmidt echo under time reversal \cite{wei2018localization}. Concretely, the double quantum Hamiltonian is given in equation \ref{eqn.DQ} below. 
\begin{equation}\label{eqn.DQ}
    \mathcal{H}_{DQ} = \frac{u}{2}\sum_{i<j} J_{ij}^{FF}\big(\hat{S}_x^{(i)} \hat{S}_x^{(j)} - \hat{S}_y^{(i)} \hat{S}_y^{(j)} \big)\\
\end{equation}
Using Wei16, the interacting scaling parameter $u$ can be varied and made negative. For these experiments, the sign of the interaction term is made negative by transforming the phases of the pulses in the Hamiltonian engineering sequence, namely via a reflection which maps $x \leftrightarrow y$ and $\overline{x} \leftrightarrow \overline{y}$. This ensures all interpulse delays are unchanged relative to the forward and backward evolution steps, leading to a more uniform error term across the entire experiment and a higher overall fidelity. Additional details of the Wei16 sequence can be found in \ref{sec.sup}. The experimental data is shown in Figure \ref{fig5}, and demonstrates a significant improvement in the Loschmidt echo decay timescale for frame-change corrected sequences of all interaction magnitudes.

\section{Conclusion}

In this work, we have demonstrated our ability to diagnose and correct for unitary error caused by short, high power, unitary control. Given the electronic origin of the phase transient error, we expect that this error modality extends beyond NMR platforms. We argued that avoiding previously utilized solutions such as pulse-shape engineering is advantageous in order to keep control pulses as short as possible. Indeed, we found better Hamiltonian engineering and time suspension performance at shorter pulse lengths, as we were able to more effectively reduce Trotter error. As expected from electronics analysis, shorter pulses require higher power, and hence lead to a larger phase transient error, which we are then able to correct using the frame-change technique.

We applied the frame-change technique to multiple Hamiltonian engineering experiments, including time suspension and time reversal. We found that the error-corrected sequences never reduce the signal fidelity, and almost always outperformed the uncorrected sequences. Since the origin of the error mode is the presence of non-linear circuit elements within the resonance probe, we conjecture that the phase transient is a prevalent error in many quantum devices with similar control architectures. Thus, the frame change technique has broad applications to devices where pulse-shaping is not feasible or desired.

\section*{Acknowledgements}
This work was supported in part by the National Science Foundation under Grants No. PHY1915218 and No. PHY1734011, and the CQE-LPS Doc Bedard Fellowship.

 \bibliographystyle{elsarticle-num} 
 \bibliography{refs}

\clearpage

\appendix

\section{Electronic Origins of Pulse Error}\label{sec.app_pulse}

It has been previously  shown with careful circuit analysis that the phase transient is due to finite rise and fall times of the electronic signal generated by the NMR spectrometer \cite{ellett1971spectrometers, barbara1991phase}.  For a simple circuit model, Ellet \textit{et. al.} showed that given an incident  electronic signal $e(t)$ and circuit impedance $Z(s)$, the current in the inductor producing the rf control pulses is given by an inverse Laplace transform of the ratio of these quantities \cite{ellett1971spectrometers}. Namely,
\begin{equation}
    i(t) = \mathcal{L}^{-1}\big[\mathcal{L}[e(t)](s)/Z(s)\big],
\end{equation}
where $\mathcal{L}$ denotes the Laplace transform, and $\mathcal{L}^{-1}$ its inverse. Via this analysis, the phase transient is a simple consequence of a non-constant electronic signal amplitude.

\begin{figure}
    \centering
    \includegraphics[width=\textwidth]{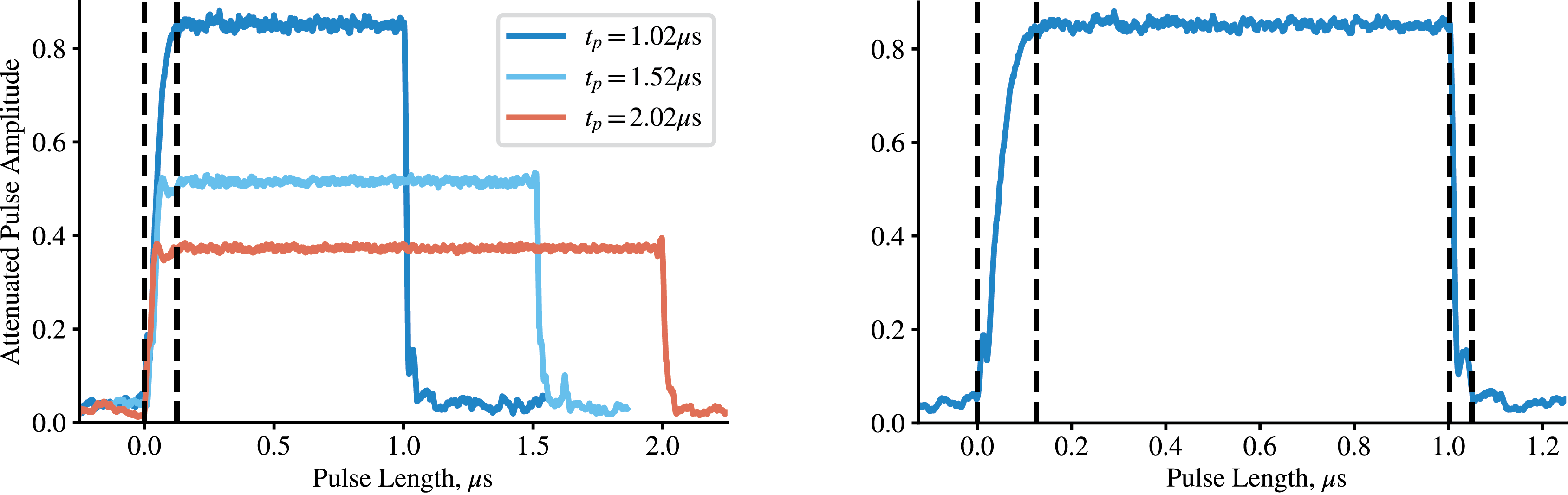}
    \caption{Oscilloscope measurements of $\pi/2$-pulses generated by the NMR spectrometer, smoothed with a simple moving average to reduce electronic noise, for $t_p\in\lbrace 1.02, 1.52, 2.02\rbrace \mu$s. On the left, we denote the start and end of the rise-time for the 1.02 $\mu$s via vertical dashed lines. On the right, we plot only the $t_p=1.02\mu$s $\pi/2$-pulse to emphasize that the rise-time portion of the pulse makes up about 12.5\% of the total duration, whereas the fall-time portion is extremely rapid by comparison.}
    \label{fig.A1}
\end{figure}

Thus, while we cannot easily use electronics to directly measure the phase transient in the probe circuit, we can still determine where a phase transient component will occur. To do this, we use an oscilloscope to measure the signal produced by the NMR spectrometer. Then, any region where the pulse envelope is non-constant (amplitude in the rotating frame) will lead to an out-of-phase component in the probe circuit. Due to technical considerations, we passed the signal through a large attenuator and long transmission line before measurement, which introduced additional sources of electronic noise. To overcome the noise we performed a simple moving average on the signal with a delay kernel of 100 data points, corresponding to a real-time window of $0.01\mu$s. Since this window is small compared to the overall pulse length, we do not expect that this averaging will meaningfully impact the pulse shape details. The results of these measurements are shown in Figure \ref{fig.A1}

 It is apparent from Figure \ref{fig.A1} that the rise-time is an increasing function of pulse power. Hence, the proportion of time spent with a time varying signal amplitude for a short high-power pulse is much larger than in a longer and lower-power pulse. This coincides with the observed trend that the shortest considered pulse has the largest phase transient error, and monotonically decreases with increasing pulse length. Further, since the rise-time is apparently significantly longer than the fall-time, the phase transient error can be well approximated as occurring only on the leading edge. This is captured by the unitary error model given in the main text in equation \ref{eqn.error_model}.

\section{Evolution under MREV-8}\label{sec.app_mrev}
The MREV-8 sequence on a solid state sample cancels the homonuclear dipolar interaction, while allowing for a residual local field \cite{haeberlen1968coherent,mansfield1971symmetrized}. Namely, the disordered local field generated by nearby phosphorus spins in fluorapatite,
\begin{equation}\label{eqn.a1}
     \mathcal{H}_{dis} = \sum_{i,j} J^{FP}_{ij} \hat{S}_z^{(i)}\hat{I}_z^{(j)} \approx \sum_{i} h_i \hat{S}_z^{(i)},
\end{equation}
is transformed to
\begin{equation}
    \mathcal{H}_{MREV} = \frac{1}{3}\sum_i h_i \big(\hat{S}_x^{(i)} + \hat{S}_z^{(i)}\big).
\end{equation}
We treat the system as having quenched disorder, since the timescale of the evolution of the phosphorus spins exceeds the experimental timescales of interest.

Thus, treating the disorder as fixed during a single experiment, we can independently compute the evolution of each spin and average over the distribution of the disorder. Then,
\begin{equation}
    \hat{S}_z^{(i)}(t) = \exp\big(-\ii t \mathcal{H}_{MREV}\big) \hat{S}_z^{(i)} \exp\big(\ii t \mathcal{H}_{MREV}\big),
\end{equation}
which can be exactly computed,
\begin{align}
    \nonumber \hat{S}_z^{(i)}(t) &= \frac{1}{2}\bigg(1+\cos\frac{\sqrt{2} h_i t}{3}\bigg)\hat{S}_z^{(i)} + \frac{1}{2}\bigg(1-\cos\frac{\sqrt{2} h_i t}{3}\bigg)\hat{S}_x^{(i)}\\
    &- \frac{\sqrt{2}}{4}\sin\frac{\sqrt{2} h_i t}{3}\hat{S}_y^{(i)}.
\end{align}
Relevant to our experiments, we can easily compute two-point infinite temperature correlator,
\begin{equation}
    4\avg{\hat{S}^{(i)}_z(t)\hat{S}^{(i)}_z} = \frac{1}{2}\bigg(1+\cos\frac{\sqrt{2} h_i t}{3}\bigg).
\end{equation}
By symmetry, we would get the same result for $\avg{\hat{S}_x(t)\hat{S}_x}$ after a repeat of the above computation.

\begin{figure*}[h!]
    \centering
    \includegraphics[scale=0.65]{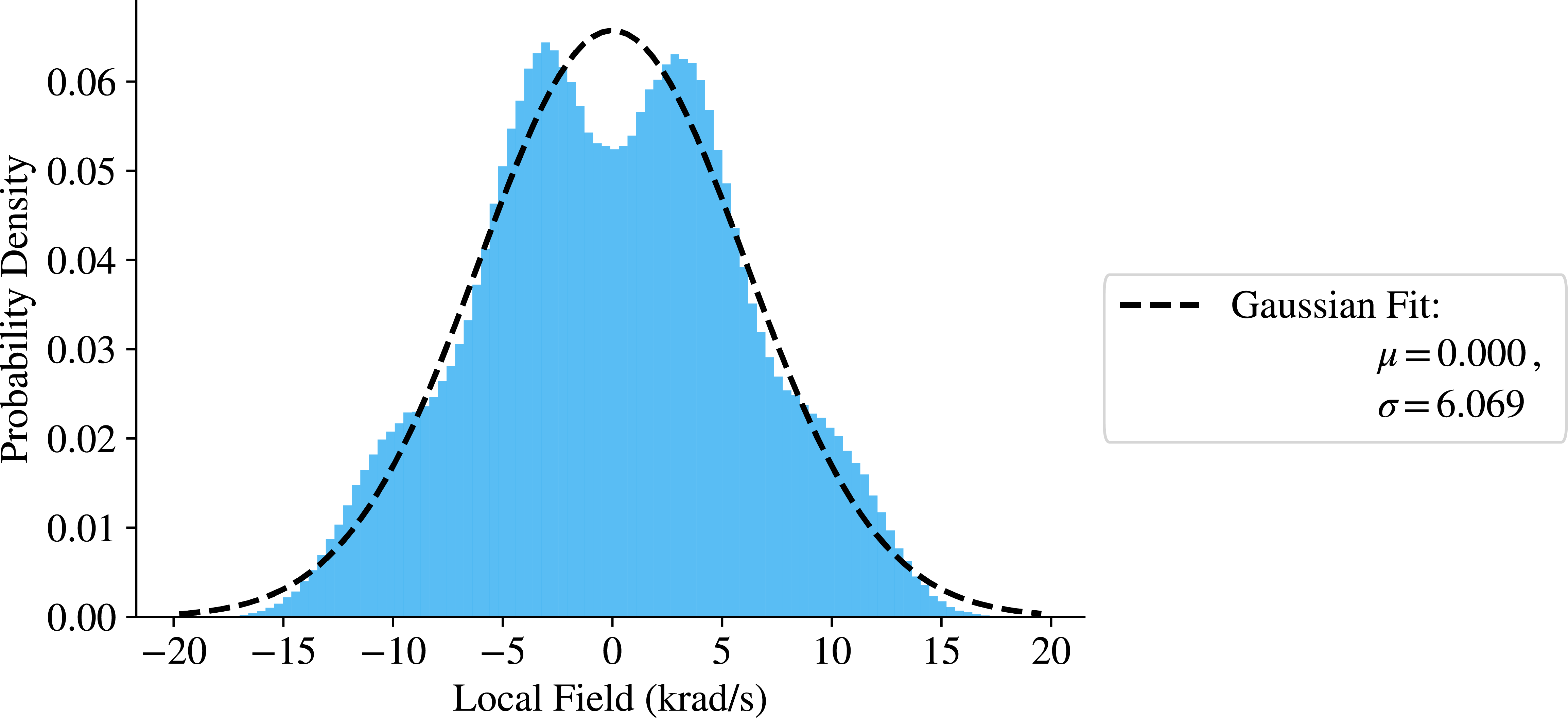}
    \caption{Disordered field distribution for a fluorine spin, as generate by neighboring phosphorus spins. The crystal is approximated as a cubic cluster of 27 neighboring unit cells containing 6 phosphorus atoms each. To avoid computing all $2^{216}$ possible configurations, we randomly sample $10^6$ configurations, where each spin is equally likely to be up or down.}
    \label{fig.A2}
\end{figure*}

Now, it remains to do the quenched disorder average. Experimentally, this happens naturally within a single scan, given that our measured signal is the average of a macroscopic number of nuclei. By Monte Carlo techniques, one can estimate the distribution of the disordered field by randomly sampling phosphorus nuclear spin orientations and computing the resulting mean field for each configuration. This technique leads to a multimodal distribution which is well approximated by a single Gaussian, with zero mean and standard deviation of $\sigma_h = 6.069$ krad s$^{-1}$, shown in figure \ref{fig.A2}. Then, averaging over the disorder distribution, we conclude that
\begin{align}
    \nonumber \overline{\avg{\hat{S}_z(t)\hat{S}_z}} &\propto \frac{1}{2}\int_{-\infty}^\infty \dd h_i \bigg(1+\cos\frac{\sqrt{2} h_i t}{3}\bigg) \frac{1}{\sqrt{2\pi}\sigma_h}e^{-\frac{h_i^2}{2 \sigma_h^2}}\\
    &= \frac{1}{2} + \frac{1}{2}e^{-\frac{1}{9}(\sigma_h t)^2}.
\end{align}
In general, if $h_i \sim X$ for arbitrary probability distribution $X$, then the Gaussian term is replaced by the characteristic function of $X$,
\begin{equation}
    \overline{\avg{\hat{S}_z(t)\hat{S}_z}} \propto \frac{1}{2} + \frac{1}{2} \text{Re}\bigg(\phi_X\big(\frac{\sqrt{2}}{3}t\big)\bigg).
\end{equation}
Here, the characteristic function is defined as $\phi_X(t) = \mathbb{E}[\exp(\ii t X)]$.

The true distribution is clearly not a Gaussian, with details that may be computationally intractable to access. However, we can still extract descriptive statistics analytically. Consider the distribution of a generalized locally disordered field, $\lbrace h_i \rbrace $. Motivated by equation (\ref{eqn.a1}), and under the assumption of quenched disorder, each spin operator $\hat{I}_z^{(j)}$ will contribute only a single eigenvalue to a given disorder instance. Hence, we are able to replace the spin operator $\hat{I}_z^{(j)}$ with a classical random variable, $I_j$. Then, the disordered field distribution is given by the formal sum,
\begin{equation}
    h_i = \sum_j J_{ij} I_j.
\end{equation}
For an infinite temperature system, $I_j$ is uniformly random with outcomes depending on the underlying nuclear spin species.

Let $s_j$ be the total spin of nucleus $j$. In a thermal ensemble consisting of a single spin species at infinite temperature, all $I_j$ are independent and identically distribution uniform random variables with outcomes given by the set of spin eigenvalues, $\lbrace -s_j, -s_{j}+1, \cdots, s_j \rbrace$. Thus, the mean is easily seen to be
\begin{equation}
    \mathbb{E}[I_j]=0,
\end{equation}
by symmetry. The second moment is given by
\begin{equation}
    \mathbb{E}[I_j^2] = \frac{1}{2s_j+1} \sum_{i=-s_j}^{s_j}{i^2} = \frac{2s_j(s_j+1)(2s_j+1)}{6(2s_j+1)} = \frac{s_j(s_j+1)}{3},
\end{equation}
which holds regardless of whether $s_j$ is integer or half-integer.

Using the above information, we can compute descriptive statistics of the disordered field variable $h_i$. The mean of the local disordered field distribution is 0 by linearity of the expectation value: 
\begin{align*}
\mu = \mathbb{E} [h_i]  &=  \mathbb{E}\left[\sum_j{J_{ij}^{FP} I_j}\right] = \sum_j{J_{ij}^{FP} \mathbb{E} [I_j]} = 0.
\end{align*}
Moreover, we can compute the variance:
\begin{align}
\nonumber (\Delta h_i)^2 &= \mathbb{E}[h_i^2] - \mathbb{E} [h_i]^2 = \mathbb{E}[h_i^2] \\
\nonumber &= \mathbb{E}\left[(\sum_j{J_{ij}^{FP} I_j})^2\right] = \mathbb{E}\left[\sum_{j,k}{J_{ij}J_{ik} I_jI_k}\right] \\
&\nonumber = \sum_{j}{J_{ij}^2 \mathbb{E}[I_j^2]} + \sum_{j\neq k}{{J_{ij}J_{ik} \mathbb{E}[I_jI_k]}}\\
&= \frac{1}{3}\sum_{j}{(J_{ij})^2 (s_j)(s_j+1)}.
\end{align}
Above, we utilized the fact that each $I_j$ is i.i.d., so that for $j\neq k$, $\mathbb{E}[I_jI_k] = \mathbb{E}[I_j]\mathbb{E}[I_k] = 0$. 

Connecting these results back to the main text, we focus on measurements of fluorapatite, such that $\hat I_z^{(j)}$ are spin operators associated with phosphorous-31 spins, hence all $s_j =\frac{1}{2}$. By translational symmetry the index $i$ can be suppressed, and we compute
\begin{equation}\label{variance}
\sigma_h^2 = (\Delta h_i)^2 = \frac{1}{4}\sum_{j}{(J_{ij}^{FP})^2}.
\end{equation}
Since the dipolar coupling strength falls off as $r^{-3}$, $J_{ij}^2 \sim \abs{i-j}^{-6}$ and will certainly converge for any 3-dimensional crystal.

Numerically, we can compute $\sigma_h$ exactly for a crystal composed of a finite number of unit cells. Then, in the infinite scaling limit with an $r^{-3}$ power law, we estimate $\sigma_h = 5.84826$ krad $s^{-1}$. From exact computations with crystal sizes on the order of $10^4$ unit cells, we find that the difference between the infinite system estimate and finite size computations is on the order of milliradians per second. These computations are qualitatively consistent with the Gaussian estimation from the Monte Carlo reconstruction of the disorder distribution.

For the case $s_j = 1/2, \forall j$, we can extend this result to finite temperatures. In this case, we have that secondary spin species is in a Gibbs state with inverse temperature $\beta$ with respect to the Zeeman Hamiltonian for a paramagnetic spin species, $\mathcal{H} = -\hbar \omega \hat{I}_z$. The first two moments are straightforward to compute:
\begin{align}
    \mathbb{E}[I_j] &= \avg{\hat{I}_z^{(j)}}_\beta = \frac{1}{2} \tanh{\frac{\beta\hbar\omega}{2}},\\
    \mathbb{E}[I_j^2] &=  \frac{1}{4}\avg{\openone}_\beta  = \frac{1}{4}.
\end{align}

Using the above single-spin expectation values, we can compute descriptive statistics of the disordered field. For non-zero $\beta$, the mean is also non-zero,
\begin{equation}
    \mu = \frac{1}{2}\tanh\frac{\beta\hbar\omega}{2} \sum_j{J_{ij}^{FP}}
\end{equation}
The variance is similarly modified,
\begin{equation}
(\Delta h_i)^2 = \sum_{j} J_{ij}^2 (\Delta I_j)^2.
\end{equation}
It's easy to see that
\begin{equation}
    (\Delta I_j)^2 = \frac{1}{4}\big(1-\tanh^2 \frac{\beta\hbar\omega}{2}\big) = \frac{1}{4}\sech^2\frac{\beta\hbar\omega}{2}.
\end{equation}
Thus,
\begin{equation}
\sigma_h^2 = \frac{1}{4}\sech^2\frac{\beta\hbar\omega}{2} \sum_{j}{(J_{ij})^2}.
\end{equation}
Indeed, for $\beta=0$, variance simplifies to the infinite temperature result. 

It is instructive to consider a small parameter expansion in $\zeta \equiv \frac{\beta\hbar\omega}{2}$. Here, $\zeta$ is unit-less, taking values of $\zeta \approx 1.5 \times 10^{-6}$ for phosphorus-31 at room temperature in a 7.1 T magnet. Hence, we can confidently employ small angle approximations $\tanh(\zeta) = \zeta + O(\zeta^3)$ and $\sech^2(\zeta) = 1 - \zeta^2 + O(\zeta^4)$. We see that our infinite temperature approximation is sufficient, as the leading order corrections are still quite small:
\begin{align*}
    \mu &= \frac{\zeta}{2}\sum_{j}{J_{ij}} + O(\zeta^3)\\
    \sigma_h^2 &= \frac{1-\zeta^2}{4}\sum_{j}{J_{ij}^2} + O(\zeta^4)
\end{align*}
Note that while at first glance $\sum_j J_{ij}$ looks to be logarithmically divergent, we argue that it converges. Upgrading to continuous spherical coordinates $(r,\theta,\phi)$, $J_{ij} \longrightarrow J(r,\theta,\phi) \propto (1-3\cos^2\theta)/r^3$ where $\theta$ is the azimuthal angle. Notice $\langle 1- 3\cos^2 \theta \rangle = 0$ on a spherical shell for uniformly distributed $\theta$. With increasing $r$, the lattice points will fill a shell, $[r, r+\delta r)$, with increasing uniformity and approach 0. As such $\sum_{j} J_{ij}$ converges quickly to a value of about $19$ krad s$^{-1}$, which we computed numerically. When multiplied by $\frac{\zeta}{2}$ this results in a disorder mean of $\mu = 0.015$ rad s$^{-1}$. In our experiments, this value is indistinguishable from 0.

\section{Supplementary Information: Hamiltonian Engineering}\label{sec.sup}

In this work, we considered the solid-state nuclear magnetic resonance of fluorapatite, Ca$_5$(PO$_4$)$_3$F, in a 300 MHz Bruker spectrometer. The crystal is placed with the magnetic field aligned along its $c$ axis, where the intrachain coupling is approximately 40 times larger than the interchain coupling~\cite{cappellaro2007simulations,zhang2009nmr,wei2018localization}. Within the rotating frame generated by the magnetic field, our system Hamiltonian has three main components: dipolar $z$, locally disordered $z$, and collective control.
\begin{equation}
    \mathcal{H}(t) = \mathcal{H}_{Dz} + \mathcal{H}_{dis} + \mathcal{H}_c(t),
\end{equation}
Concretely, each component of the Hamiltonian is:
\begin{align*}
    \mathcal{H}_{Dz} &= \frac{1}{2}\sum_{i<j} J_{ij}^{FF}\big(2\hat{S}_z^{(i)}\hat{S}_z^{(j)} - (\hat{S}_x^{(i)} \hat{S}_x^{(j)} + \hat{S}_y^{(i)} \hat{S}_y^{(j)})\big)\\
    \mathcal{H}_{dis} &= \sum_{i,j} J^{FP}_{ij} \hat{S}_z^{(i)}\hat{I}_z^{(i)} \approx \sum_{i} h_i \hat{S}_z^{(i)} \\
    \mathcal{H}_c(t) &= f(t) \big(\cos(\phi) \hat{S}_x + \sin(\phi)\hat{S}_y\big),
\end{align*}
where $J_{ij}^{FF} = J_0^{FF}/\abs{i-j}^3$ is the homonuclear fluorine dipolar coupling strength ($J_0^{FF} \approx -33 $ krad s$^{-1}$), and $J_{ij}^{FP} = (1-3\cos^2\theta_{ij}) J_0^{FP}/r_{ij}^3$ is the heteronuclear fluorine--phosphorus dipolar coupling strength. Above, we have introduced the collective spin operators, 
\begin{equation}
    \hat{S}_\mu = \sum_i \sigma_\mu^{(i)},\,\,\, \mu \in \lbrace x, y, z\rbrace.
\end{equation}
As discussed in the main text, $\mathcal{H}_c(t)$ is idealized, wherein $f(t)$ is a square pulse at fixed power so that the net effect of each control pulse is a $\pi/2$ rotation about a fixed axis determined by $\phi$. In our system, $\phi$ has a resolution of one degree, and is fixed during the duration of a pulse. 

Within the main text, we utilized a variety of multi-pulse sequences, including Wei16, Angle12, Peng24, and MREV-8. Here, we provide the pulse sequences and resulting time averaged Hamiltonians for all experiments considered.

To begin, Wei16 is a 16-pulse sequence, composed of four similarly structured 4-pulse blocks \cite{wei2018exploring}. Unless otherwise specified, a pulse corresponds to a $\pi/2$ rotation about a particular transverse axis of the Bloch sphere obtained by turning on the rf driving for a time $t_p$, which in this work takes values $t_p \in \lbrace 1.02, 1.52, 2.02 \rbrace \mu$s. Each 4-pulse block corresponds to 6$\tau_0$ of time, where $\tau_0 = 5 \mu$s is the delay primitive for the entire sequence. It is prudent to choose $\tau_0$ to be as short as possible in order to minimize Trotter errors while still ensuring interpulse delays $\tau_0-t_p$ are not too short. For our spectrometer, $\tau_0 = 5\mu$s is as short as feasible, given that the electronics require a minimum pulse separation of $2.5\mu$s to change the phase for the next applied pulse. For case where $t_p = 2.02\mu$s, $\tau_0$ must be extended to $6 \mu$s.

Under Wei16, the internal Hamiltonian can be transformed to a model with 6 free parameters, 3 controlling the interaction Hamiltonian and 3 controlling the orientation of the local field. Notably, with collective rotations we cannot change the sum of the coefficients of the interaction Hamiltonian, since the subspace generated by the dipolar interaction along $x$, $y$, and $z$, written $\text{span}\lbrace \mathcal{H}_{D_x}, \mathcal{H}_{D_y}, \mathcal{H}_{D_z}\rbrace$, is a two-dimensional vector space. This can be seen clearly by recalling one of the central identities used to perform time suspension, $\mathcal{H}_{D_x} + \mathcal{H}_{D_y} + \mathcal{H}_{D_z} = 0$. The constraint is directly captured in the Floquet Hamiltonian, defined below.
\begin{align}\label{eqn.wei16}
    \nonumber \mathcal{H}_F = \frac{1}{2}\sum_{i<j} J_{ij} \big( (u-w)&\hat{S}_x^{(i)}\hat{S}_x^{(j)} + (v-u)\hat{S}_y^{(i)}\hat{S}_y^{(j)} + (w-v)\hat{S}_z^{(i)}\hat{S}_z^{(j)}\big)\\
    &+ \frac{1}{3}\sum_i \omega_i\big(a\hat{S}_x^{(i)}+ b\hat{S}_y^{(i)} +c\hat{S}_z^{(i)}\big).
\end{align}
Each of the parameters appearing above are combined to produced the physical pulse delays for the sequence, given below.
\begin{align*}
    \tau_1 &= \tau_0(1+c-v+w), \hspace{0.5cm} \tau_2 = \tau_0(1+b-u+v),\hspace{0.5cm} \tau_3 = \tau_0(1-a+u-w)\\
    \tau_1' &= \tau_0(1-c-v+w),\hspace{0.5cm} \tau_2' = \tau_0(1-b-u+v),\hspace{0.5cm}\tau_3 = \tau_0(1+a+u-w).
\end{align*}

In general, a four-pulse sequence is defined by five delays and four orientations, which we collect using the following notation, $P(t_1, \bm{n}_1, t_2, \bm{n}_2, t_3, \bm{n}_3, t_4, \bm{n}_4, t_5)$; a standard notation for our group \cite{wei2018exploring,peng2022disorder}. Unlike strings of unitary operators, this notation is read left-to-right, in the order they are physically applied during an experiment. Namely, the generic pulse sequence corresponds to ``wait for $t_1$, then apply a pulse in the $\bm{n}_1$ direction, then wait for $t_2$, ...'' and so on. Concretely, the Wei16 sequence is given by the string of four pulse blocks shown below.
\begin{align*}
    P(\tau_1,\bm{x},\tau_2,\bm{y}, 2\tau_3,\bm{y}, \tau_2',\bm{x},\tau_1') P(\tau_1',\bm{x},\tau_2,\bm{y}, 2\tau_3',\bm{y}, \tau_2',\bm{x},\tau_1) \cdots\\
    P(\tau_1,\bar{\bm{x}},\tau_2',\bar{\bm{y}}, 2\tau_3',\bar{\bm{y}}, \tau_2,\bar{\bm{x}},\tau_1') P(\tau_1',\bar{\bm{x}},\tau_2',\bar{\bm{y}}, 2\tau_3,\bar{\bm{y}}, \tau_2,\bar{\bm{x}},\tau_1)
\end{align*}
Here we use the standard notation $\bar{\bm{x}} = -\bm{x}$, which helps save horizontal space. In this sequence, there is a great deal of reflection symmetry. Generally, symmetrized pulse sequences perform better than their unsymmetrized counterparts, which can be formalized by computing leading-order error term.

The Peng24 sequence differs from Wei16 in that it is constructed by symmetrizing a pulse sequence reliant on a three-fold symmetry, which differs significantly from the usual two-fold symmetry four-pulse blocks demonstrated above. Additionally, as a time suspension sequence, the target Hamiltonian is much simpler, $\mathcal{H}_F = 0$. In particular, we chose Peng24 as it is a 24-pulse sequence corresponding to $24\tau_0$ of evolution time, equivalent to that of Wei16, allowing for a direct comparison. We can write Peng24 in the same way we did Wei16, taking the standard pulse block to include three orientations and four delays in order to better emphasize the symmetry group. Since Peng24 is symmetrized Angle12, we define Angle12 below and make concrete how it is used to define Peng24 (first introduced as yxx24) \cite{peng2022deep}.
\begin{align*}
    \text{Angle12} = P\big(\frac{\tau_0}{2},\bar{\bm{y}},\tau_0,\bm{x},\tau_0,\bar{\bm{x}},\frac{\tau_0}{2}\big) P\big(\frac{\tau_0}{2},\bm{y},\tau_0,\bar{\bm{x}},\tau_0,\bar{\bm{x}},\frac{\tau_0}{2}\big)\cdots\\
    P\big(\frac{\tau_0}{2},\bar{\bm{y}},\tau_0,\bm{x},\tau_0,\bar{\bm{x}},\frac{\tau_0}{2}\big) P\big(\frac{\tau_0}{2},\bm{y},\tau_0,\bm{x},\tau_0,\bm{x},\frac{\tau_0}{2}\big)
\end{align*}
As stated, Peng24 is formed by symmetrizing Angle12, which is done by performing Angle12, followed by $\overline{\text{Angle12}}$. The overbar on Angle12 means to perform Angle12, albeit with the signs of each pulse orientation reversed, $\bm{x}\longrightarrow\bar{\bm{x}}$ etc. Hence, 
\begin{equation}
    \text{Peng24} = \text{Angle12}\,\,\overline{\text{Angle12}}.
\end{equation}

Finally, we review the 8-pulse sequence MREV-8 given by the following series of pulses~\cite{Mansfield73}:
\begin{equation}
    \text{MREV-8} =  P\big(\tau_0,\bar{\bm{x}},\tau_0,\bm{y},2\tau_0,\bar{\bm{y}},\tau_0,\bm{x},\tau_0\big) P\big(\tau_0,\bm{x},\tau_0,\bm{y},2\tau_0,\bar{\bm{y}},\tau_0,\bar{\bm{x}},\tau_0\big).
\end{equation}
In addition to decoupling the dipolar interaction of the fluorine atoms, MREV-8 alters the magnitude and direction of the disordered field, with a resulting Floquet Hamiltonian of
\begin{equation}
    \mathcal{H}_{MREV} =\frac{1}{3}\sum_{i} h_i \bigg(\hat{S}_x^{(i)} + \hat{S}_z^{(i)}\bigg).
\end{equation}
Other 8-pulse spectral sequences can produce different orientations of the disordered field, such as WAHUHA-8\cite{Waugh68}, which leaves the disordered field orientation unchanged, albeit similarly re-scaled in magnitude.

\end{document}